\documentclass[conference]{IEEEtran}

\usepackage{caption}
\usepackage{subcaption}

\usepackage{comment}
\usepackage{amssymb}
\usepackage{amsmath}
\usepackage{amsfonts}
\usepackage{mathrsfs}
\usepackage{engord}
\usepackage[dvips]{graphicx}
\usepackage{threeparttable}
\usepackage{booktabs}
\usepackage{epsfig}
\usepackage{color}
\usepackage{graphicx}
\usepackage{multirow}
\usepackage{cite}
\usepackage{epstopdf}

\usepackage{framed}
\usepackage{url}

\usepackage{color}

\newtheorem{thm}{Theorem}

\newtheorem{lem}{Lemma}

\newtheorem{ex}{Example}
\newtheorem{cor}{Corollary}

\newtheorem{rmk}{Remark}

\begin{document}

%
\title{Tree Structured Synthesis of Gaussian Trees}

\author{\IEEEauthorblockN{Ali Moharrer,
Shuangqing Wei,
George T. Amariucai, 
and Jing Deng}}
\maketitle
\footnotetext[1]{A. Moharrer, and S. Wei are with the school of EECS,  Louisiana State University, Baton Rouge, LA 70803, USA (Email: amohar2@lsu.edu, swei@lsu.edu). 
G. T. Amariucai is with the department of ECE, Iowa State University, Ames, IA, USA (Email: gamari@iastate.edu). 
J. Deng is with the department of CS, University of North Carolina at Greensboro, Greensboro, NC, USA (Email: jing.deng@uncg.edu).
This material is based upon work supported in part by the National Science
Foundation under Grant No. 1320351.}



\begin{abstract}
A new synthesis scheme is proposed to effectively generate a random
vector with prescribed joint density that induces a (latent) Gaussian
tree structure. The quality of synthesis is measured by 
total variation distance between the synthesized and desired statistics.
The proposed layered and successive encoding scheme relies on the learned
structure of tree to use minimal number of common random variables
to synthesize the desired density. We characterize the achievable rate
region for the rate tuples of multi-layer latent Gaussian tree, through
which the number of bits needed to simulate such Gaussian joint density
are determined.  The random sources used in our algorithm are the latent
variables at the top layer of tree, the additive independent Gaussian
noises, and the Bernoulli sign inputs that capture the ambiguity of correlation
signs between the variables.  
\end{abstract}

\section{Introduction}

Consider the problem of simulating a random vector with prescribed
joint density.  Such method can be used for prediction applications,
i.e., given a set of inputs we may want to compute the output response
statistics.  This can be achieved by generating an appropriate number
of random input bits to a stochastic channel whose output vector has its
empirical statistics meeting the desired one measured by a given metric.

We aim to address such synthesis problem for a case where the prescribed
output statistics induces a (latent) \textit{Gaussian tree} structure,
i.e., the underlying structure is a tree and the joint density of the
variables is captured by a Gaussian density.  The Gaussian graphical
models are widely studied in the literature. They have diverse  applications
in social  networks,   biology,  and  economics \cite{gauss3,gauss5},
to  name  a  few.  Gaussian  trees  in  particular have attracted much
attention \cite{gauss5} due to their sparse structures, as well as
existing computationally efficient algorithms in learning the underlying
topologies  \cite{mit,correlation}.  In this paper we assume that the
parameters and structure information of the latent Gaussian tree is
provided.

Our primary concern in such synthesis problem is about efficiency in terms of the amount of random bits required at the input, as well as the modeling complexity of given stochastic system through which the Gaussian vector is synthesized.
We use \textit{Wyner's common information} \cite{wyner} to quantify the information theoretic complexity of our scheme.
Such quantity defines the necessary number of common random bits to generate two correlated outputs, through a single common source of randomness, and two independent channels.

In \cite{verdu}, Han and Verdu introduced the notion of
\textit{resolvability} of a given channel, which is defined as the
minimal required randomness to generate output statistics in terms
of a vanishing total variation distance between the synthesized and
prescribed joint densities.  
In \cite{isit2014,trans2016,cuff_common}, the authors aim to define the common
information of $n$ dependent random variables, to further address the
same question in this setting.  

In this paper, we show that unlike previous cases, the Gaussian trees can be synthesized not only using a single variable as a common source, but by relying on vectors (usually consisting of more than one variable).
In particular, we consider an input vector (and not a single variable) to produce common
random bits, and adopt a specific (but natural) structure
to our synthesis scheme to decrease the number of parameters needed to model the
synthesis scheme.  It is worthy to point that the achievability results
given in this paper are under the assumed structured encoding framework.
Hence, although through defining an optimization problems, we show that
the proposed method is efficient in terms of both modeling and codebook
rates, the converse proof, which shows the optimality of such scheme
and rate regions is postponed to future studies.

We show that in latent Gaussian trees, we are always having a sign singularity issue \cite{arxiv_version}, which we can exploit to make our synthesis approach more efficient.
To clarify, consider the following example.

\begin{ex}
Consider a Gaussian tree shown in Figure \ref{fig:broadcast}.
It consists of three observed variables $X_1$, $X_2$, and $X_3$ that are connected to each other through a single hidden node $Y^{(1)}$.
Define \small$\rho_{x_iy}=E[X_iY]$ \normalsize as the true correlation values between the input and each of the three output.
Define $B^{(1)}\in\{-1,1\}$ \normalsize as a binary variable that as we will see reflects the sign information of pairwise correlations. 
For the star structure shown in Figure \ref{fig:broadcast}, one may assume that $B^{(1)}=1$ to show the case with $\rho'_{x_iy}=\rho_{x_iy}$, while $B^{(1)}=-1$ captures $\rho''_{x_iy}=-\rho_{x_iy}$, where $\rho'_{x_iy}$ and $\rho''_{x_iy},~i\in\{1,2,3\}$, are the recovered correlation values using certain inference algorithm such as RG \cite{mit}.
It is easy to see that both recovered correlation values induce the same covariance matrix $\Sigma_{\mathbf{x}}$, showing the sign singularity issue in such a latent Gaussian tree.
In particular, for each pairwise correlation $\rho_{x_ix_j}$, we have $\rho_{x_ix_j}=\rho_{x_iy}\rho_{x_jy}=(B^{(1)})^2\rho_{x_iy}\rho_{x_jy}$, where the second equality is due to the fact that regardless of the sign value, the term $(B^{(1)})^2$ is equal to $1$.
Now, depending on whether we replace $B^{(1)}$ with $\{1,-1\}$, we obtain $\rho_{x_ix_j}=\rho'_{x_iy}\rho'_{x_jy}=\rho''_{x_iy}\rho''_{x_jy}$.
And there is no way to distinguish these two groups using only the given information on observables joint distribution.
\begin{figure} [h]
\centering 
\includegraphics[width=0.25\columnwidth]{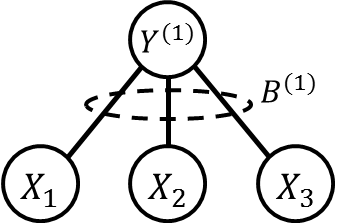}
\caption{A simple Gaussian tree with a hidden node $Y^{(1)}$\label{fig:broadcast}} 
\end{figure}
\end{ex}


It turns out that such sign singularity can be seen as another \textit{noisy} source of randomness, which can further help us to reduce the code-rate corresponding to latent inputs to synthesize the latent Gaussian tree.
In fact, we may think of the Gaussian tree shown in Figure \ref{fig:broadcast} as a communication channel, where information flows from a Gaussian source \small $Y^{(1)}\sim N(0,1)$ \normalsize through three communication channels $p_{X_i|Y^{(1)}}(x_i|y^{(1)})$ with independent additive Gaussian noise variables \small $Z_i\sim N(0,\sigma^2_{z_i}),~i\in\{1,2,3\}$ \normalsize to generate (dependent) outputs with \small $\mathbf{X}\sim N(0,\Sigma_{\mathbf{x}})$. 
\normalsize We introduce \small $B^{(1)}\in\{-1,1\}$ \normalsize as a binary Bernoulli random variable, which reflects the sign information of pairwise correlations.
Our goal is to characterize the achievable rate region and through an encoding scheme to synthesize Gaussian outputs with density $q_\mathbf{X}(\mathbf{x})$ using only Gaussian inputs and through a channel with additive Gaussian noise, where the synthesized joint density $q_\mathbf{X}(\mathbf{x})$ is indistinguishable from the true output density $p_\mathbf{X}(\mathbf{x})$ as measured by \textit{total variation} metric \cite{cuff}.

\section{Problem Formulation} \label{sec:formulation}

\subsection{The signal model of a multi-layer latent Gaussian tree}
Here, we suppose a latent graphical model, with $\mathbf{Y}=[Y_1,Y_2,...,Y_k]'$ as the set of inputs (hidden variables), $\mathbf{B}=[B_1,...,B_m]$, with each $B_i\in\{-1,1\}$ being a binary Bernoulli random variable with parameter $\pi_i=p(B_i=1)$ to introduce sign variables, and $\mathbf{X}=[X_1,X_2,...,X_n]'$ as the set of Gaussian outputs (observed variables) with $p_{\mathbf{X}}(\mathbf{x})$.
We also assume that the underlying network structure is a latent Gaussian tree, therefore, making the joint probability (under each sign realization) $p_{\mathbf{XY|B}}$ be a Gaussian joint density $N(\mathbf{\mu},\Sigma_{\mathbf{xy|b}})$, where the covariance matrix $\Sigma_{\mathbf{xy|b}}$ induces tree structure $G_T(V,E,W)$, where $V$ is the set of nodes consisting of both vectors $\mathbf{X}$ and $\mathbf{Y}$; $E$ is the set of edges; and $W$ is the set of edge-weights determining the pairwise covariances between any adjacent nodes.
We consider normalized variances for all variables $X_i\in \mathbf{X},~i\in\{1,2,...,n\}$ and $Y_j\in\mathbf{Y},~j\in\{1,2,...,k\}$. Such constraints do not affect the tree structure, and hence the independence relations captured by $\Sigma_{\mathbf{xy|b}}$.
Without loss of generality, we also assume $\mathbf{\mu}=\mathbf{0}$, this constraint does not change the amount of information carried by the observed vector. 

In \cite{arxiv_version} we showed that the vectors $\mathbf{X}$ and $\mathbf{B}$ are independent.
We argued the intrinsic sign singularity in Gaussian trees is due to the fact that the pairwise correlations $\rho_{x_ix_j}\in \Sigma_\mathbf{x}$ can be written as $\prod_{(l,k)\in E} \rho_{x_lx_k}$, i.e., the product of correlations on the path from $x_i$ to $x_j$.
Hence, roughly speaking, one can carefully change the sign of several correlations of the path, and still maintain the same value for $\rho_{x_ix_j}$.
We showed that if the cardinality of the input vector $\mathbf{Y}$ is $k$, then $2^k$ minimal Gaussian trees (that only differ in sign of pairwise correlations) may induce the same joint Gaussian density $p_{\mathbf{X}}$ \cite{arxiv_version}.

In order to propose the successive synthesis scheme, we need to characterize the definition of \textit{layers} in a latent Gaussian tree.
We define latent vector $\mathbf{Y}^{(l)}$, to be at layer $l$, if the shortest path between each latent input $Y_i^{(l)}\in\mathbf{Y}^{(l)}$ and the observed layer (consisting the output vector $\mathbf{X}$) is through $l$ edges.
In other words, beginning from a given latent Gaussian tree, we assume the output to be at layer $l=0$, then we find its immediate latent inputs and define $\mathbf{Y}^{(1)}$ to include all of them.
We iterate such procedure till we included all the latent nodes up to layer $L$, i.e., the top layer.
In such setting, the sign input vector $\mathbf{B}^{(l)}$ with Bernoulli sign random variables $B_i^{(l)}\in\mathbf{B}^{(l)}$ is assigned to the latent inputs $\mathbf{Y}^{(l)}$.

We adopt a communication channel to feature the relationship between each pair of successive layers.
Assume $\mathbf{Y}^{(l+1)}$ and $\mathbf{B}^{(l+1)}$ as the input vectors, $\mathbf{Y}^{(l)}$ as the output vector, and the noisy channel to be characterized by the conditional probability distribution $P_{\mathbf{Y}^{(l)}|\mathbf{Y}^{(l+1)},\mathbf{B}^{(l+1)}}(\mathbf{y}^{(l)}|\mathbf{y}^{(l+1)},\mathbf{b}^{(l+1)})$, the signal model for such a channel can be written as follows,
\begin{align} \label{eq:multi-layer_linear_regression}
\mathbf{Y}^{(l)}=\mathbf{A_B}^{(l+1)}\mathbf{Y}^{(l+1)}+\mathbf{Z}^{(l+1)},
~~~l\in[0,L-1]
\end{align}
\noindent \noindent where $\mathbf{A_B}^{(l+1)}$ is the $|\mathbf{Y}^{(l)}|\times |\mathbf{Y}^{(l+1)}|$ transition matrix that also carries the sign information vector $\mathbf{B}^{(l+1)}$, and $\mathbf{Z}^{(l+1)}\sim N(0,\Sigma_{\mathbf{z}^{(l)}})$ is the additive Gaussian noise vector with independent elements, each corresponding to a different communication link from the input layer $l+1$ to the output layer $l$.
Hence, the outputs $\mathbf{Y}^{(l)}$ at each layer $l$, are generated using the inputs $\mathbf{Y}^{(l+1)}$ at the upper layer.
The case $l=0$, is essentially for the outputs in $\mathbf{X}$, which will be produced using their upper layer inputs at $\mathbf{Y}^{(1)}$.
As we will see next, such modeling will be the basis for our successive encoding scheme.
In fact, by starting from the top layer inputs $L$, at each step we generate the outputs at the lower layer, this will be done till we reach the observed layer to synthesize the Gaussian vector $\mathbf{X}$.
Finally, note that in order to take all possible latent tree structures, we need to revise the ordering of layers in certain situations, which will be taken care of in their corresponding subsections.
For now, the basic definition for layers will be satisfactory.

\subsection{Synthesis Approach Formulation}
In this section we provide mathematical formulations to address the following fundamental problem: using channel inputs $\mathbf{Y}^{(l+1)}$ and $\mathbf{B}^{(l+1)}$, what are the rate conditions under which we can synthesize the Gaussian channel output $\mathbf{Y}^{(l)}$ with a given $p_{\mathbf{Y}^{(l)}|\mathbf{B}^{(l)}}$, for each $l\in [0,L-1]$.
Note that, at first we are only given $p_{\mathbf{X}}$, but using certain tree learning algorithms we can find those jointly Gaussian latent variables $p_{\mathbf{Y}^{(l)}|\mathbf{B}^{(l)}}$ at every level $l\in [1,L]$.
We propose a successive encoding scheme on multiple layers that together induce a latent Gaussian tree, as well as the corresponding bounds on achievable rate tuples.
The encoding scheme is efficient because it utilizes the latent Gaussian tree structure to simulate the output.
In particular, without resorting to such learned structure we need to characterize $\mathcal{O}(kn)$ parameters (one for each link between a latent and output node), while by considering the sparsity reflected in a tree we only need to consider $\mathcal{O}(k+n-1)$ parameters (the edges of a tree).

Suppose we transmit input messages through $N$ channel uses, in which $t\in\{1,2,...,N\}$ denotes the time index.
We define $\vec{Y}^{(l)}_{t}[i]$ to be the $t$-th symbol of the $i$-th codeword, with $i\in C_{\mathbf{Y}^{(l)}}=\{1,2,...,M_{Y^{(l)}}\}$ where $M_{Y^{(l)}}=2^{NR_{Y^{(l)}}}$ is the codebook cardinality, transmitted from the existing $k_l$ sources at layer $l$.
We assume there are $k_l$ sources $Y^{(l)}_j$ present at the $l$-th layer, and the channel has $L$ layers.
We can similarly define $\vec{B}^{(l)}_{t}[k]$ to be the $t$-th symbol of the $k$-th codeword, with $k\in C_{\mathbf{B}^{(l)}}=\{1,2,...,M_{B^{(l)}}\}$ where $M_{B^{(l)}}=2^{NR_{B^{(l)}}}$ is the codebook cardinality, regarding the sign variables at layer $l$.
We will further explain that although we define codewords for the Bernoulli sign vectors as well, they are not in fact transmitted through the channel, and rather act as \textit{noisy sources} to select a particular sign setting for latent vector distributions.
For \textit{sufficiently} large rates $R_{\mathbf{Y}}=[R_{Y^{(1)}},R_{Y^{(2)}},...,R_{Y^{(L)}}]$ and $R_{\mathbf{B}}=[R_{B^{(1)}},R_{B^{(2)}},...,R_{B^{(L)}}]$ and as $N$ grows the synthesized density of latent Gaussian tree converges to $p_{\mathbf{W}^N(\mathbf{w}^N)}$, i.e., $N$ i.i.d realization of the given output density $p_{\mathbf{W}}(\mathbf{w})$, where $W=\{\mathbf{X,Y,B}\}$ is a compound random variable consisting the output, latent, and sign variables.
In other words, the average total variation between the two joint densities vanishes as $N$ grows \cite{cuff},
\begin{align} \label{eq:TV}
\lim_{N\rightarrow\infty} E||q(\mathbf{w}_1,...,\mathbf{w}_N)-\prod_{t=1}^N p_{\mathbf{w}_t}(\mathbf{w}_t)||_{TV}\rightarrow 0
\end{align}
\noindent where $q(\mathbf{w}_1,...,\mathbf{w}_N)$ is the synthesized density of latent Gaussian tree, and $E||.||_{TV}$, represents the average total variation.
In this situation, we say that the rates $(R_{\mathbf{Y}},R_{\mathbf{B}})$ are \textit{achievable} \cite{cuff}.
Our achievability proofs heavily relies on \textit{soft covering lemma} shown in \cite{cuff}.

Loosely speaking, the soft covering lemma states that one can synthesize the desired statistics with arbitrary accuracy, if the codebook size (characterized by its corresponding rate) is sufficient and the channel through which these codewords are sent is noisy enough.
This way, one can \textit{cover} the desired statistics up to arbitrary accuracy, hence, any random sample that can be drawn from the desired distribution $p_{\mathbf{w}}$, it also exists in the synthesized distribution $q_{\mathbf{w}}$.
The main objective is to maximize such rate region (hence minimizing the required codebook size), and develop a proper encoding scheme to synthesize the desired statistics.

For simplicity of notation, we drop
the symbol index and use $Y^{(l)}_{t}$ and $B^{(l)}_{t}$ instead of
$\vec{Y}^{(l)}_{t}[i]$ and $\vec{B}^{(l)}_{t}[k]$, respectively, since
they can be understood from the context.

\section{Main Results}

Based on the proposed layered model for a Gaussian tree, we always end up with two cases: $1)$ those cases where the variables $Y_i^{(l)}\in \mathbf{Y}^{(l)}$ at the same layer $l$ are not adjacent to each other, for any $l\in [0,L]$; $2)$ those cases where the variables at the same layer $l$ can be adjacent.
An example for the first case in shown in Figure \ref{fig:B2}, where there is no edge between the variables in the same layer of a two layered Gaussian tree.
Also, Figure \ref{fig:internal2} shows a Gaussian tree capturing the second case.
As we discuss, the synthesis scheme for each of these cases is different.
In particular, in the second case we need to pre-process the Gaussian tree and change the ordering of variables at each layer, and then perform the synthesis.

\subsection{The case with observables at the same layer}
\begin{figure} [h!]
\centering 
\includegraphics[width=0.9\columnwidth]{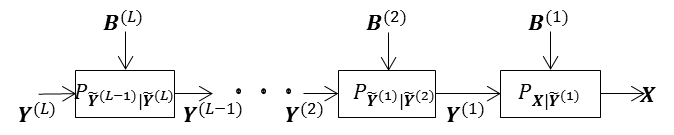}
\caption{Multi-layered output synthesis\label{fig:layered}} 
\end{figure}

In this case, Figure \ref{fig:layered} shows the general encoding scheme to be used to synthesize the output vector.

In particular, to synthesize the output joint distribution $p_{\mathbf{Y}^{(l)}}$ at each layer $l$, we need to generate two codebooks $C_{\mathbf{Y}^{(l+1)}}$ and $C_{\mathbf{B}^{(l+1)}}$ at its upper layer $l+1$.
Then, we need to follow certain synthesis scheme to send such codewords on each of these channels to generate the entire Gaussian tree statistics.
To better clarify our approach, it is best to begin the synthesis discussion by an illustrative example.

\begin{ex} \label{ex:multi_layer}
Consider the case shown in Figure \ref{fig:B2}, in which the Gaussian tree has two layers of inputs.
\begin{figure} [h!]
\centering 
\includegraphics[width=0.7\columnwidth]{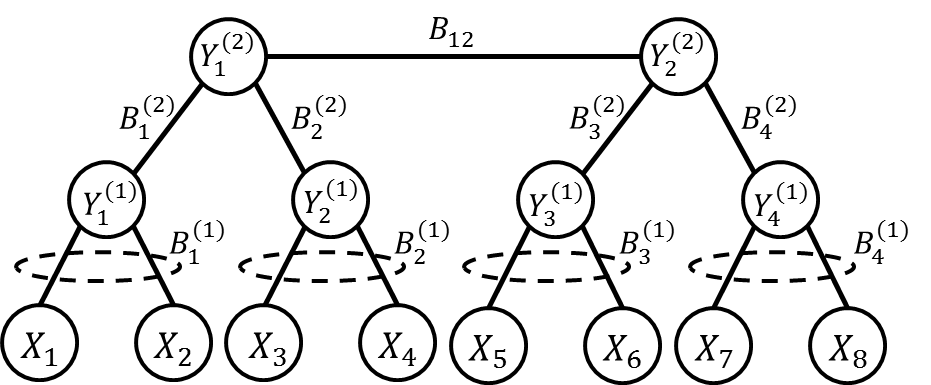}
\caption{Two-layered Gaussian Tree\label{fig:B2}} 
\end{figure} 
We can write the pairwise covariance between inputs at the first layer as $E[Y_{k,t}^{(1)}Y_{l,t}^{(1)}|\mathbf{B}^{(1)}]=\gamma_{kl}B_{k,t}^{(1)}B_{l,t}^{(1)}$, in which $k\neq l\in\{1,2,3,4\}$. 
We know that the input vector $\mathbf{Y}_{t}^{(1)}$ becomes Gaussian for each realization of $\mathbf{B}_{t}^{(1)}=\{\mathbf{b}_{1,t}^{(1)},\mathbf{b}_{2,t}^{(1)},\mathbf{b}_{3,t}^{(1)},\mathbf{b}_{4,t}^{(1)}\}$. 
Hence, one may divide the codebook $\mathbb{C}$ into at most $2^4=16$ parts $\mathbb{S}_i,~i\in\{1,2,...,16\}$, in which each part follows a specific Gaussian density with covariance values $E[Y_{k,t}^{(1)}Y_{l,t}^{(1)}|\mathbf{b}^{(1)}]=\gamma_{kl}b_{k,t}^{(1)}b_{l,t}^{(1)},~k\neq l\in\{1,2,3,4\}$.
Then, the lower bound on the possible rates in the second layer is as follows, 
\begin{align} \label{eq:rates_fig_4b}
&R_{\mathbf{Y}^{(2)}}\geq I(\mathbf{Y}^{(1)};\mathbf{Y}^{(2)}|\mathbf{B}^{(1)})\notag\\
&R_{\mathbf{Y}^{(2)}}+R_{\mathbf{B}^{(2)}}\geq I(\mathbf{Y}^{(1)};\mathbf{Y}^{(2)},\mathbf{B}^{(2)}|\mathbf{B}^{(1)})
\end{align}
The formal results on general cases will be given in Theorem \ref{thm:achievability_basic}.
Let us elaborate the synthesis scheme in this case, which will serve as a foundation for our successive encoding scheme proposed for any general Gaussian tree.

First, we need to generate the codebooks at each layer, beginning from the top layer all the way to the first layer.
The sign codebooks $C_{\mathbf{B}^{(2)}}$ and $C_{\mathbf{B}^{(1)}}$ are generated beforehand, and simply regarding the Bernoulli distributed sign vectors $\mathbf{B}^{(2)}$, and $\mathbf{B}^{(1)}$.
Hence, each sign codeword is a sequence of vectors consisting elements chosen from $\{-1,1\}$.
We may also generate the top layer codebook $C_{\mathbf{Y}^{(2)}}$ using mixture Gaussian codewords, where each codeword at each time slot consists of all possible sign realizations of $\mathbf{B}_t^{(2)}$.

Each of these settings characterize a particular Gaussian distribution for the top layer latent variables $\mathbf{Y}^{(2)}$.
The necessary number of codewords needed is $M_{\mathbf{Y}^{(2)}}=2^{NR_{\mathbf{Y}^{(2)}}}$, where the rate in the exponent is lower bounded and characterized using \eqref{eq:rates_fig_4b}.
To form the second codebook $C_{\mathbf{Y}^{(1)}}$, we know that we should use the codewords in $C_{\mathbf{Y}^{(2)}}$.
We randomly pick codewords from $C_{\mathbf{Y}^{(2)}}$ and $C_{\mathbf{B}^{(2)}}$.
Now, based on the chosen sign codeword, we form the sequence $(\mathbf{y}^{(2)}|\mathbf{b}^{(2)})^N$ to be sent through the channels.
The number of channels is determined by the sign vector $\mathbf{B}^{(1)}$.
In this example, $\mathbf{B}^{(1)}$ consists of $k_1=4$ sign variables, hence, resulting in $2^{k_1}=16$ different channel realizations.
Hence, by sending the chosen codeword $(\mathbf{y}^{(2)}|\mathbf{b}^{(2)})^N$ through these $16$ channels, we produce a particular codeword $(\mathbf{y}^{(1)})^N$ for the first layer codebook $C_{\mathbf{Y}^{(1)}}$.
Note that, such produced codeword is in fact a collection of Gaussian vectors, each corresponding to a particular sign realization $\mathbf{b}^{(1)}\in\mathbf{B}^{(1)}$.
We iterate this procedure $M_{\mathbf{Y}^{(1)}}$ times to produce enough codewords that are needed for synthesis requirements of the next layer.
The necessary size of $M_{\mathbf{Y}^{(1)}}$ is lower bounded by Theorem \ref{thm:achievability_basic}.

Figure \ref{fig:encoding_general}, shows the described synthesis procedure.
In order to produce an output sequence, all we need to do is to randomly pick codewords from $C_{\mathbf{Y}^{(1)}}$ and $C_{\mathbf{B}^{(1)}}$.
Then, depending on each time slot sign realization $\mathbf{B}_t^{(1)}$ we use the corresponding channel $p_{\mathbf{X}|\mathbf{Y}^{(1)}\mathbf{B}_t^{(1)}}$ to generate a particular output sequence $\mathbf{X}^N$.
Note that the middle \textit{bin} between the two codebooks in Figure \ref{fig:encoding_general} is not a codebook.
It is used to show the sufficiency of the size of $C_{\mathbf{Y}^{(2)}}$ codebook, and the noisy enough channel $p_{\mathbf{Y}^{(1)}|\mathbf{Y}^{(2)}\mathbf{B}^{(2)}\mathbf{b}^{(1)}}$ to cover the desired joint density of $p_{\mathbf{Y}^{(1)}}$ up to arbitrary accuracy.
\begin{figure} [h!]
\centering 
\includegraphics[width=0.9\columnwidth]{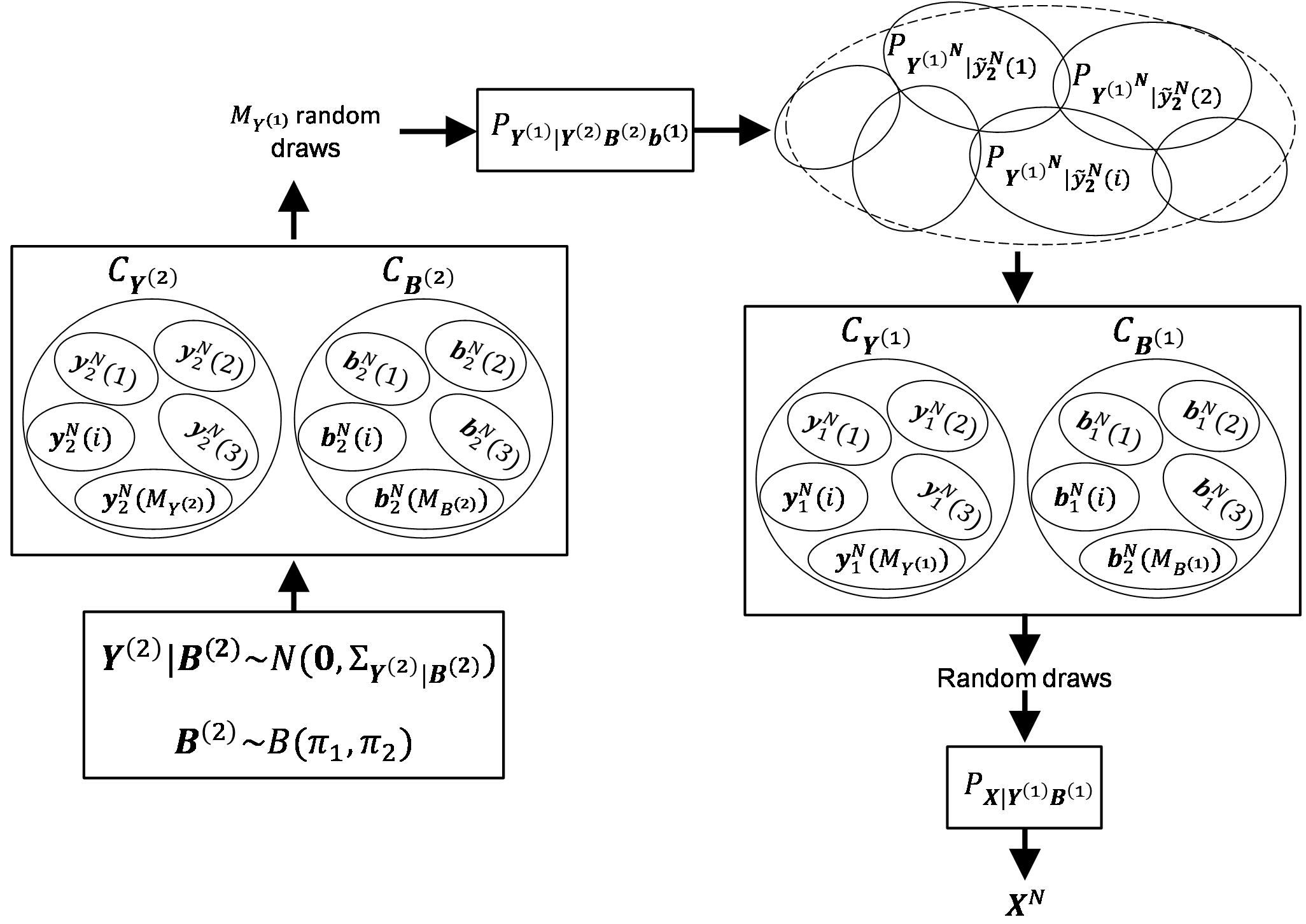}
\caption{The proposed encoding scheme used for a Gaussian tree shown in Figure \ref{fig:B2}.}
\label{fig:encoding_general}
\end{figure}


In general, the output at the $l$-th layer $\mathbf{Y}^{(l)}$ is synthesized by $\mathbf{Y}^{(l+1)}$ and $\mathbf{B}^{(l+1)}$, which are at layer $l+1$.
This is done through iterating the successive encoding scheme described above.
In particular, looking from the bottom layer (output layer), the vector $\mathbf{X}$ is synthesized using the upper layer inputs $\mathbf{Y}^{(1)}$ and through the channel characterized by first layer sign variables $\mathbf{B}^{(1)}$.
Now, such inputs themselves are generated using their upper layer variables $\mathbf{Y}^{(2)}$ and through the channels regarding $\mathbf{B}^{(2)}$.
This procedure is continued till we reach the top layer nodes.
The only variables who are generated independently and using a random number generator, are the top layer latent variables $\mathbf{Y}^{(L)}$.
Therefore, we only need independent Gaussian noises at each layer as a source of randomness to gradually synthesize the output that is close enough to the true observable output, measured by total variation.
In Theorem \ref{thm:achievability_basic}, whose proof can be found in Appendix \ref{app:achievability_basic} we obtain the achievable rate region for multi-layered latent Gaussian tree, while taking care of sign information as well, i.e., at each layer dividing a codebook into appropriate sub-blocks capturing each realization of sign inputs.
\begin{thm} \label{thm:achievability_basic}
{\it
For a latent Gaussian tree having $L$ layers, and forming a hyper-chain structure, the achievable rate region is characterized by the following inequalities for each layer $l$,
\begin{align} \label{eq:thm_achievability_basic}
&R_{\mathbf{B}^{(l+1)}}+R_{\mathbf{Y}^{(l+1)}} \geq I[\mathbf{Y}^{(l+1)},\mathbf{B}^{(l+1)};\mathbf{Y}^{(l)}|\mathbf{B}^{(l)}]\notag\\
&R_{\mathbf{Y}^{(l+1)}}\geq I[\mathbf{Y}^{(l+1)};\mathbf{Y}^{(l)}|\mathbf{B}^{(l)}],~l\in[0,L-1]
\end{align}
}
\end{thm}
\noindent where $l=0$ shows the observable layer, in which there is no conditioning needed, since the output vector $\mathbf{X}$ is already assumed to be Gaussian.

\end{ex}

\subsection{The case with observables at different layers}
\begin{figure} [h!]
\centering
\includegraphics[width=0.3\columnwidth]{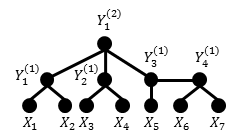} 
\caption{Latent Gaussian tree with adjacent nodes at layer $1$}\label{fig:internal2} 
\end{figure}

To address this case, we need to reform the latent Gaussian tree structure by choosing an appropriate root such that the variables in the newly introduced layers mimic the basic scenario, i.e., having no edges between the variables at the same layer.
We begin with the top layer nodes and as we move to lower layers we seek each layer for the adjacent nodes at the same layer, and move them to a newly added layer in between the upper and lower layers.
This way, we introduce new layers consisting of those special nodes, but this time we are dealing with a basic case.
Note that such procedure might place the output variables at different layers, i.e., all the output variables are not generated using inputs at a single layer.
We only need to show that using such procedure and previously defined achievable rates, one can still simulate output statistics with vanishing total variation distance.
To clarify, consider the case shown in Figure \ref{fig:internal2}.

\begin{figure} [h!]
\centering
\includegraphics[width=0.3\columnwidth]{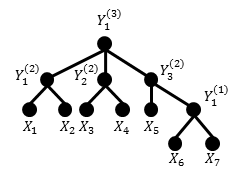} 
\caption{Another layer introduced to address the issue}\label{fig:internal3} 
\end{figure}

As it can be seen, there are two adjacent nodes in the first layer, i.e., $Y_3^{(1)}$ and $Y_4^{(1)}$ are connected.
Using the explained procedure, we may move $Y_4^{(1)}$ to another newly introduced layer, then we relabel the nodes again to capture the layer orderings.
The reformed Gaussian tree is shown in Figure \ref{fig:internal3}.
In the new ordering, the output variables $X_6$ and $X_7$ will be synthesized one step after other outputs.
The input $Y_1^{(3)}$ is used to synthesize the vector $\mathbf{Y}^{(2)}$, and such vector is used to generate the first layer outputs, i.e., $X_1$ to $X_5$ and $Y_1^{(1)}$.
At the last step, the input $Y_1^{(1)}$ will be used to simulate the output pair $X_6$ and $X_7$.
By Theorem \ref{thm:achievability_basic} we know that both simulated densities regarding to $q_{X^N_1,X^N_2,X^N_3,X^N_4,X^N_5}$ and $q_{X^N_6,X^N_7}$ approach to their corresponding densities as $N$ grows.
We need to show that the overall simulated density $q_{\mathbf{X}^N}$ also approaches to $\prod_{t=1}^N p_{\mathbf{X}}(\mathbf{x}_t)$ as well.

We need to be extra cautious in keeping the joint dependency among the generated outputs at different layers: For each pair of outputs $(X^N_6,X^N_7)$, there exists an input codeword $(Y_1^{1})^N$, which corresponds to the set of generated codewords $(X^N_1,X^N_2,X^N_3,X^N_4,X^N_5)$, where together with $(Y_1^{1})^N$ they are generated using the second layer inputs.
Hence, in order to maintain the overall joint dependency of the outputs, we always need to match the correct set of outputs $X_1^N$ to $X_5^N$ to each of the output pairs $(X^N_6,X^N_7)$, where this is done via $(Y_1^{1})^N$.

In general, we need to keep track of the indices of generated output vectors at each layer and match them with corresponding output vector indices at other layers.
This is shown in Lemma \ref{lem:achievability_general2}, whose proof can be found in Appendix \ref{app:achievability_general2}.
\begin{lem} \label{lem:achievability_general2}
{\it
For a latent Gaussian tree having $L$ layers, by rearranging each layer so that there is no intra-layer edges, the achievable rate region at each layer $l$ is characterized by the same inequalities as in Theorem \ref{thm:achievability_basic}.
}
\end{lem}

\section{Maximum Achievable Rate Regions under Gaussian Tree Assumption}

Here, we aim to minimize the bounds on achievable rates shown in \eqref{eq:thm_achievability_basic} to make our encoding more efficient.
Considering the first lower bound in \eqref{eq:thm_achievability_basic}, we derived an interesting result in \cite{arxiv_version} that shows for any Gaussian tree such mutual information value is only a function of given $p_{\mathbf{X}}(\mathbf{x})\sim N(\mathbf{0},\Sigma_{\mathbf{x}})$.
However, considering the second second inequality in \eqref{eq:thm_achievability_basic}, in Theorem
\ref{thm:uniform_sign}, whose proof can be found in Appendix \ref{app:uniform_sign}.
we show that under Gaussian tree assumption, the lower bound $I[\mathbf{Y}^{(l+1)};\mathbf{Y}^{(l)}|\mathbf{B}^{(l)}],~l\in[0,L-1]$
is minimized for homogeneous Bernoulli sign inputs.
\begin{thm} \label{thm:uniform_sign}
{\it
Given the Gaussian vector $\mathbf{X}$ with $\Sigma_{\mathbf{x}}$
inducing a latent Gaussian tree, with latent parameters
$\mathbf{Y}$ and sign vector $\mathbf{B}$ the optimal
solution for $\mathbf{\pi}^*=arg\min_{\mathbf{\pi}^{(l)}\in [0,1]^{|\mathbf{B}^{(l)}|}}
I[\mathbf{Y}^{(l+1)};\mathbf{Y}^{(l)}|\mathbf{B}^{(l)}],~l\in[0,L-1]$ is at $\pi^{(l)}_i = p[B_i^{(l)}=1]=1/2$ for all $B_i^{(l)}\in \mathbf{B}^{(l)}$ and at each layer $l\in [1,L]$}.
\end{thm}

\section{Conclusion} \label{sec:conclusion}
In this paper, we proposed a new tree structure synthesis scheme,
in which through layered forwarding channels certain Gaussian vectors
can be efficiently generated.  Our layered encoding approach is shown
to be efficient and accurate in terms of reduced required number of
parameters and random bits needed to simulate the output statistics, and its closeness
to the desired statistics in terms of total variation distance.

\bibliography{reference}
\bibliographystyle{IEEEtran}

\appendices

\section{Proof of Theorem \ref{thm:uniform_sign}} \label{app:uniform_sign}

Suppose the latent Gaussian tree has $k$ latent variables,i.e., $\mathbf{Y}=[Y_1,Y_2,...,Y_k]$.
By adding back the sign variables the joint density $p_{\mathbf{XY}}$ becomes a Gaussian mixture model.
One may model such mixture as the summation of densities that are conditionally Gaussian, given sign vector.
\begin{align} \label{eq:f_i}
p_{\mathbf{XY}}(\mathbf{x,y}) = \sum_{i=0}^{2^k-1} \eta_{\mathbf{B}_i}f_i(\mathbf{x,y})
\end{align}

\noindent where each $\eta_{\mathbf{B}_i}$ captures the overall probability of the binary vector $\mathbf{B}_i=[b_{1i},b_{2i},...,b_{ki}]$, with $b_{ji}\in\{0,1\}$.
Here, $b_{ji}=0$ is equivalent to having $b_{ji}=-1$.
The terms $f_i(\mathbf{x,y})$ are conditional densities of the form $p(\mathbf{x,y}|\mathbf{B}_i)$

In order to characterize $I(\mathbf{X},\mathbf{Y})$, we need to find $p_{\mathbf{Y}}(\mathbf{y})$ in terms of $\eta_{\mathbf{B}_i}$ and conditional Gaussian densities as well.
First, let's show that for any two hidden nodes $y_i$ and $y_j$ in a latent Gaussian tree, we have $E[y_iy_j]=\rho_{y_iy_j}b_ib_j$. The proof goes by induction: 
We may consider the structure shown in Figure \ref{fig:B2} as a base, where we proved that $B_{12}=B^{(1)}_1B^{(1)}_2$.
Then, assuming such result holds for any Gaussian tree with $k-1$ hidden nodes, we prove it also holds for any Gaussian tree with $k$ hidden nodes. Let's name the newly added hidden node as $y_k$ that is connected to several hidden and/or observable such that the total structure forms a tree. 
Now, for each newly added edge we assign $b_kb_{n_k}$, where $n_k\in N_k$ is one of the neighbors of $y_k$. Note that this assignment maintains the pairwise sign values between all previous nodes,
since to find their pairwise correlations we go through $y_k$ at most once, where upon entering/exiting $y_k$ we multiply the correlation value by $b_k$, hence producing $b_k.b_k=1$, so overall the pairwise correlation sign does not change. 
Note that the other pairwise correlation signs that do not pass through $C_k$ remain unaltered.
One may easily check that by assigning $b_kb_{n_k}$ to the sign value of each newly added edge we make $y_k$ to follow the general rule, as well.
Hence, overall we showed that $E[y_iy_j]=\rho_{y_iy_j}b_ib_j$ for any $y_i,y_j\in\mathbf{Y}$.
This way we may write $\Sigma_{\mathbf{y}}=B\Sigma'_{\mathbf{y}}B$, where $\rho_{y_iy_j}\in\Sigma'_{\mathbf{y}}$ and $b_i\in B$ is $k\times k$ diagonal matrix.
One may easily see that both $B$ and its negation matrix $-B$ induce the same covariance matrix $\Sigma_\mathbf{y}$. As a result,  if we define $\eta_{\bar{\mathbf{B}}_i}$ as a compliment of $\eta_{\mathbf{B}_i}$, we can write the mixture density $p_{\mathbf{Y}}(\mathbf{y})$ as follows,
\begin{align} \label{eq:g_i}
p_{\mathbf{Y}}(\mathbf{y}) = \sum_{i=0}^{2^{k-1}-1} (\eta_{\mathbf{B}_i}+\eta_{\bar{\mathbf{B}}_i})g_i(\mathbf{y})
\end{align}

\noindent where the conditional densities can be characterized as $g_i(\mathbf{y})=p(\mathbf{y}|\mathbf{B}_i)=p(\mathbf{y}|\bar{\mathbf{B}}_i)$.
We know that $g_i(\mathbf{y})=\int f_j(\mathbf{x,y}) d\mathbf{x}$, where $j$ may correspond to either $\mathbf{B}_i$ or $\bar{\mathbf{B}}_i$.

First, we need to show that the mutual information $I(\mathbf{X},\mathbf{Y})$ is a convex function of $\eta_{\mathbf{B}_i}$ for all $i\in[0,2^k-1]$. By equality $I(\mathbf{X},\mathbf{Y})=h(\mathbf{X})-h(\mathbf{X}|\mathbf{Y})$, and knowing that given $\Sigma_{\mathbf{x}}$ the entropy $h(\mathbf{X})=1/2\log (2\pi e)^n|\Sigma_{\mathbf{x}}|$ is fixed, we only need to show that the conditional entropy $h(\mathbf{X}|\mathbf{Y})$ is a concave function of $\eta_{\mathbf{B}_i}$.
Using definition of entropy and by replacing for $p_{\mathbf{XY}}$ and $p_{\mathbf{Y}}$ using equations \eqref{eq:f_i} and \eqref{eq:g_i}, respectively, we may characterize the conditional entropy.
By taking second order derivative, we deduce the following,
\begin{align} \label{eq:second_derivative}
\dfrac{\partial^2 h(\mathbf{X}|\mathbf{Y})}{\partial^2\eta_i\eta_j} =& -\int\int\dfrac{f_i(\mathbf{x,y})f_j(\mathbf{x,y})}{p_{\mathbf{XY}}}d\mathbf{x}d\mathbf{y} \notag\\&+ \int\dfrac{\tilde{g}_i(\mathbf{y})\tilde{g}_j(\mathbf{y})}{p_{\mathbf{Y}}}d\mathbf{y}
\end{align}

\noindent where for simplicity of notations we write $\eta_i$ instead of $\eta_{\mathbf{B}_i}$. Also, $\tilde{g}_i(\mathbf{y})=\tilde{g}_{\bar{i}}(\mathbf{y})=g_i(\mathbf{y})$ for $i\in[0,2^{k-1}-1]$.
Note the following relation,
\begin{align} \label{eq:equality1}
\int\int\dfrac{\tilde{g}_i(\mathbf{y})f_j(\mathbf{x,y})p_{\mathbf{X|Y}}}{p_{\mathbf{XY}}}d\mathbf{x}d\mathbf{y} &= \int\int\dfrac{\tilde{g}_i(\mathbf{y})f_j(\mathbf{x,y})}{p_{\mathbf{Y}}}d\mathbf{x}d\mathbf{y}\notag\\
&=\int\dfrac{\tilde{g}_i(\mathbf{y})}{p_{\mathbf{Y}}}(f_j(\mathbf{x,y})d\mathbf{x})d\mathbf{y}\notag\\
&=\int\dfrac{\tilde{g}_i(\mathbf{y})\tilde{g}_j(\mathbf{y})}{p_{\mathbf{Y}}}d\mathbf{y}
\end{align}

The same procedure can be used to show, 
\begin{align} \label{eq:equality2}
\int\int\dfrac{\tilde{g}_j(\mathbf{y})f_i(\mathbf{x,y})p_{\mathbf{X|Y}}}{p_{\mathbf{XY}}}d\mathbf{x}d\mathbf{y}=\int\dfrac{\tilde{g}_i(\mathbf{y})\tilde{g}_j(\mathbf{y})}{p_{\mathbf{Y}}}d\mathbf{y}
\end{align}

By equalities shown in \eqref{eq:equality1} and \eqref{eq:equality2}, it is straightforward that \eqref{eq:second_derivative} can be turn into the following,
\begin{align}
h_{ij}=\dfrac{\partial^2 h(\mathbf{X}|\mathbf{Y})}{\partial^2\eta_i\eta_j} = -&\int\int\dfrac{1}{p_{\mathbf{XY}}}
[f_i(\mathbf{x,y})-\tilde{g}_i(\mathbf{y})p_{\mathbf{X|Y}}]\times\notag\\
&[f_j(\mathbf{x,y})-\tilde{g}_j(\mathbf{y})p_{\mathbf{X|Y}}]d\mathbf{x}d\mathbf{y}
\end{align}

The matrix $H=[h_{ij}],~i,j\in[0,2^k-1]$ characterizes the Hessian matrix the conditional entropy $h(\mathbf{X|Y})$. To prove the concavity, we need to show $H$ is non-positive definite.
Define a non-zero real row vector $\mathbf{c}\in R^{2^k}$, then we need to form $\mathbf{c}H\mathbf{c}^T$ as follows and show that it is non-positive.
\begin{align}
\mathbf{c}H\mathbf{c}^T=-\int\int\dfrac{1}{p_{\mathbf{XY}}}&
\sum_{i=0}^{2^k-1}\sum_{j=0}^{2^k-1} c_ic_j[f_i(\mathbf{x,y})-\tilde{g}_i(\mathbf{y})p_{\mathbf{X|Y}}]\notag\\
&[f_j(\mathbf{x,y})-\tilde{g}_j(\mathbf{y})p_{\mathbf{X|Y}}]d\mathbf{x}d\mathbf{y}\notag\\
=-\int\int\dfrac{1}{p_{\mathbf{XY}}}
[&\sum_{i=0}^{2^k-1}c_i(f_i(\mathbf{x,y})-\tilde{g}_i(\mathbf{y})p_{\mathbf{X|Y}})]^2d\mathbf{x}d\mathbf{y}\notag\\
&\leq 0
\end{align}

Now that we showed the concavity of the conditional entropy with respect to $\eta_i$, we only need to find the optimal solution. The formulation is defined in \eqref{eq:lagrange}, where $\lambda$ is the Lagrange multiplier.
\begin{align} \label{eq:lagrange}
L = h(\mathbf{X}|\mathbf{Y}) - \lambda\sum_{i=0}^{2^k-1} \eta_i
\end{align}

\noindent by taking derivative with respect to $\eta_i$, we may deduce the following,
\begin{align}
\dfrac{\partial L}{\partial \eta_i} = &-\int\int f_i(\mathbf{x,y})\log p_{\mathbf{XY}}d\mathbf{x}d\mathbf{y}\notag\\
&+ \int \tilde{g}_i(\mathbf{y})\log p_{\mathbf{Y}}d\mathbf{y} - \lambda\notag\\
&=-\int\int f_i(\mathbf{x,y})\log p_{\mathbf{X|Y}}d\mathbf{x}d\mathbf{y} - \lambda
\end{align}

\noindent where the last equality is due to $\tilde{g}_i(\mathbf{y})=\int f_i(\mathbf{x,y})d\mathbf{x}$.
One may find the optimal solution by solving $\partial L/\partial\eta_i=0$ for all $i\in[0,2^k-1]$, which results in showing that $-\int\int [f_i(\mathbf{x,y})-f_j(\mathbf{x,y})]\log p_{\mathbf{X|Y}}d\mathbf{x}d\mathbf{y}=0$, for all $i,j\in[0,2^k-1]$.
In order to find the joint Gaussian density $f_i(\mathbf{x,y})$, observe that we should compute the exponent $[\mathbf{xy}]\Sigma^{-1}_{\mathbf{xy}}[\mathbf{xy}]'$. Since, we are dealing with a latent Gaussian tree, the structure of $U=\Sigma^{-1}_{\mathbf{xy}}$ can be summarized into four blocks as follows \cite{chernoff}. $U_\mathbf{x}$ that has diagonal and off-diagonal entries $u_{x_i}$ and $u_{x_ix_j}$, respectively, and not depending on the edge-signs;
$U_{\mathbf{xy}}$, with nonzero elements $u_{x_iy_j}$ showing the edges between $x_i$ and particular $y_j$ and depending on correlation signs;
$[U_{\mathbf{xy}}]^T$;
$U_{\mathbf{y}}$, with nonzero off diagonal elements $u_{y_iy_j}$  that are a function of edge-sign values, while the diagonal elements $u_{y_i}$ are independent of edge-sign values.
One may show,
\begin{align}
[\mathbf{xy}]\Sigma^{-1}_{\mathbf{xy}}[\mathbf{xy}]'
&= [\sum_{i=1}^n x_i^2 u_{x_i} + \sum_{i=1}^k y_i^2 u_{y_i}]\notag\\
&+ 2[\sum_{n^x_{y_1}} x_iy_1u_{x_iy_1}+...+\sum_{n^x_{y_k}} x_iy_ku_{x_iy_k}]\notag\\
&+ 2\sum_{(i,j)\in E_X} x_ix_ju_{x_ix_j} + 2\sum_{(i,j)\in E_Y} y_iy_ju_{y_iy_j}\notag\\
&= t + 2\sum_{i=1}^k p_i + 2s + 2\sum_{(i,j)\in E_Y} y_iy_ju_{y_iy_j}
\end{align}

\noindent where $n^x_{y_i}$ are the observed neighbors of $y_i$, and $E_Y$ is the edge set corresponding only to hidden nodes, i.e., those hidden nodes that are adjacent to each other.
$E_X$ can be defined similarly, with $s=\sum_{(i,j)\in E_X} x_ix_ju_{x_ix_j}$. 
Also $p_j=\sum_{n^x_{y_j}} x_iy_ju_{x_iy_j}$.
Suppose $f_i(\mathbf{x,y})$ and $f_j(\mathbf{x,y})$ are different at $l$ sign values $\{i_1,...,i_l\}\in L$. Let's write,
\begin{align}
\sum_{(i,j)\in E_Y} y_iy_ju_{y_iy_j}& = \sum_{\substack{{(i,j)\in E_Y}\\{{i,j\in L}}~or~{i,j\notin L}}} y_iy_ju_{y_iy_j}\notag\\
&+ \sum_{\substack{{(i,j)\in E_Y}\\{i~or~j\in L}}} y_iy_ju_{y_iy_j}\notag\\
& = q + q'
\end{align}

Hence, we divide the summation $\sum_{(i,j)\in E_Y} y_iy_ju_{y_iy_j}$ into two parts $q$ and $q'$.
Suppose $\eta_i=1/2^k$ for all $i\in[0,2^k-1]$.
We may form $f_i(\mathbf{x,y})-f_j(\mathbf{x,y})$ as follows,
\begin{align*}
f_i(\mathbf{x,y})-f_j(\mathbf{x,y})\propto &e^{-t/2+s+q+\sum_{i\notin L} p_i}\notag\\
&\times[e^{q'+\sum_{i\in L} p_i}-e^{-q'-\sum_{i\in L} p_i}]
\end{align*}

By negating all $y_{i_1},...,y_{i_l}$ into $-y_{i_1},...,-y_{i_l}$, it is apparent that $t$, $\sum_{i\notin L} p_i$, and $s$ do not change. Also, the terms in $q$ either remain intact or doubly negated, hence, overall $q$ remains intact also. 
However, by definition, $p_i,i\in L$ will be negated, hence overall the sum $\sum_{i\in L} p_i$ will be negated. The same thing can be argued for $q'$, since exactly one variable $y_i$ or $y_j$ in the summation, will change its sign, so $q'$ also will be negated.
Overall, we can see that by negating $y_{i_1},...,y_{i_l}$, we will negate $f_i-f_j$.
It remains to show that such negation does not impact $p_{\mathbf{X|Y}}$. Note that since $p_{\mathbf{XY}}$ includes all $2^k$ sign combinations and all of $f_i(\mathbf{x,y})$ are equi-probable since we assumed $\eta_i=1/2^k$ so $p_{\mathbf{XY}}$ is symmetric with respect to $\eta_i$, and such transformation on $y_{i_1},...,y_{i_l}$ does not impact the value of $p_{\mathbf{XY}}$, since by such negation we simply switch the position of certain Gaussian terms $f_i(\mathbf{x,y})$ with each other.

For $p_{\mathbf{y}}$, we should first compute the term $\mathbf{y}\Sigma_{\mathbf{y}}^{-1}\mathbf{y}'$. We know $\Sigma_{\mathbf{y}}=B\Sigma'_{\mathbf{y}} B$, so $\Sigma^{-1}_Y=B^{-1}\Sigma'^{-1}_{\mathbf{y}} B^{-1}=B\Sigma'^{-1}_{\mathbf{y}} B$ (note, $\Sigma_{\mathbf{y}}$ does not necessarily induce a tree structure). 
We have,
\begin{align*}
\mathbf{y}\Sigma_{\mathbf{y}}^{-1}\mathbf{y}'= \sum_{i=1}^k w_{ii} y_i^2 + 2\sum_{i,j\& i<j} w_{ij} y_iy_jb_ib_j 
\end{align*}

From this equation, we may interpret the negation of $y_{i_1},...,y_{i_l}$, simply as negation of $b_{i_1},...,b_{i_l}$. Hence, since $p_{\mathbf{y}}$ includes all sign combinations, hence, such transformation only permute the terms $\tilde{g}_i(\mathbf{y})$, so $p_{\mathbf{y}}$ remains fixed.
Hence, overall $p_{\mathbf{X|Y}}$ remains unaltered.
As a result, we show that for any given point in the integral $\int\int (f_i(\mathbf{x,y})-f_j(\mathbf{x,y}))\log p_{\mathbf{X|Y}}d\mathbf{x}d\mathbf{y}$ we can find its negation, hence making the integrand an odd function, and the corresponding integral zero.
Hence, making the solution $\eta_i=1/2^k$, for all $i\in[0,2^k-1]$ an optimal solution.

The only thing remaining is to show that from $\eta_i=1/2^k$ we may conclude that $\pi_j=1/2$ for all $j\in[1,k]$.
By definition, we may write,
\begin{align*}
\eta_i=\prod_{j=1}^k \pi_j^{b_{ji}}(1-\pi_j)^{1-b_{ji}}
\end{align*}

\noindent where $b_{ji}\in B_i$.
Assume all $\eta_i=1/2^k$. Consider $\eta_1$ and find $\eta_{i^*}$ such that the two are different in only one expression, say at the $l$-th place.
Since, all $\eta_i$ are equal, one may deduce $1-\pi_l=\pi_l$ so $\pi_l=1/2$. Note that such $\eta_{i^*}$ can always be found since $\eta_i$'s are covering all possible combinations of $k$-bit vector. 
Now, find another $\eta_{j^*}$, which is different from $\eta_1$ at some other spot, say $l'$, again using similar arguments, we may show $\pi_{l'}=1/2$.
This can be done $k$ times to show that, if all $\eta_i=1/2^k$, then $\pi_1=...=\pi_k=1/2$.
This completes the proof.
%
%
%
%

\section{Proof of Theorem \ref{thm:achievability_basic}} \label{app:achievability_basic}
The signal model can be directly written as follows,
\begin{align} \label{new_signal}
\mathbf{Y}^{(l)} = A_{\mathbf{B}^{(l+1)}} \mathbf{Y}^{(l+1)} + \mathbf{Z}^{(l)}
\end{align}

\begin{figure} [h!]
\centering
\includegraphics[width=0.75\columnwidth]{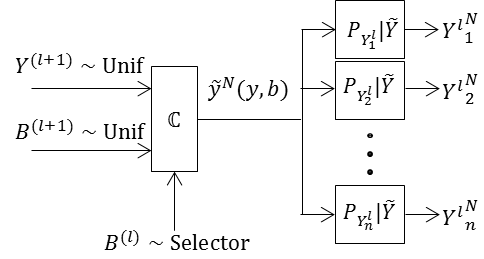} 
\caption{Construction of the joint density $\gamma_{(\mathbf{Y}^{(l)})^N,\mathbf{Y}^{(l+1)},\mathbf{B}^{(l+1)}}$}\label{fig:encoding_basic} 
\end{figure}

Here, we show the encoding scheme to generate $\mathbf{Y}^{(l)}$ from $\mathbf{Y}^{(l+1)}$.
Note that $\mathbf{Y}^{(l)}$ is a vector consisting of the variables $Y_i^{(l)}$. Also, $\mathbf{Y}^{(l+1)}$ is a vector consisting of variables $Y_i^{(l+1)}$.
The proof relies on the procedure taken in \cite{cuff}. Note that our encoding scheme should satisfy the following constraints,
\begin{tabular}{l l}
$1) ({Y}^{(l)}_i)^N\perp ({Y}^{(l)}_j)^N|\tilde{\mathbf{Y}}^{(l+1)}~~ (i\neq j)$\\
$2) (\mathbf{Y}^{(l)})^N\perp \mathbf{B}^{(l+1)}$\\
$3) P_{(\mathbf{Y}^{(l)})^N} = \prod_{t=1}^N P_{\mathbf{Y}^{(l)}}(\mathbf{y}^{(l)}_t)$\\ 
$4) |\mathbf{Y}^{(l+1)}|= 2^{NR_{\mathbf{Y}^{(l+1)}}}$\\
$5) |\mathbf{B}^{(l+1)}|= 2^{NR_{\mathbf{B}^{(l+1)}}}$\\
$6) ||q_{(\mathbf{Y}^{(l)})^N}-\prod_{t=1}^N P_{\mathbf{Y}^{(l)}}(\mathbf{y}^{(l)}_t)||_{TV}<\epsilon$
\end{tabular}

\noindent where the first constraint is due to the conditional independence assumption characterized in the signal model \eqref{new_signal}. The second one is to capture the intrinsic ambiguity of the latent Gaussian tree to capture the sign information. Condition $3)$ is due to independence of joint densities $P_{\mathbf{Y}^l}(\mathbf{Y}^l_t)$ at each time slot $t$. Conditions $4)$ and $5)$ are due to corresponding rates for each of the inputs $\mathbf{Y}^{(l+1)}$ and $\mathbf{B}^{(l+1)}$. And finally, condition $6)$ is the synthesis requirement to be satisfied.
First, we generate a codebook $\mathcal{C}$ of $\tilde{y}^N$ sequences, with indices $y\in C_Y=\{1,2,...,2^{NR_{\mathbf{Y}^{(l+1)}}}\}$ and $b\in C_B=\{1,2,...,2^{NR_{\mathbf{B}^{(l+1)}}}\}$ according to the explained procedure explained in Example \ref{ex:multi_layer}.
The codebook $\mathcal{C}$ consists of all combinations of the sign and latent variables codewords, i.e., $|\mathcal{C}| = |C_Y|\times |C_B|$.
We construct the joint density $\gamma_{(\mathbf{Y}^{(l)})^N,\mathbf{Y}^{(l+1)},\mathbf{B}^{(l+1)}}$ as depicted by Figure \ref{fig:encoding_basic}.

The indices $y$ and $b$ are chosen independently and uniformly from the codebook $\mathcal{C}$.
As can be seen from Figure \ref{fig:encoding_basic}, for each $\mathbf{B}^{(l)}_t = \mathbf{b}^{(l)}_t$ the channel $P_{{Y}^l|\tilde{Y}}$ is in fact consists of $n$ independent channels $P_{{Y}^l_i|\tilde{Y}},~i\in\{1,2,...,n\}$. The joint density is as follows,
\begin{align*}
\gamma_{(\mathbf{Y}^{(l)})^N,\mathbf{Y}^{(l+1)},\mathbf{B}^{(l+1)}}=\dfrac{1}{|C_Y||C_B|}[\prod_{t=1}^N P_{\mathbf{Y}^l}(\mathbf{Y}^l_t|\tilde{y}_t(y,b))]
\end{align*}

Note that $\gamma_{(\mathbf{Y}^{(l)})^N,\mathbf{Y}^{(l+1)},\mathbf{B}^{(l+1)}}$ already satisfies the constraints $1)$, $4)$, and $5)$ by construction.
Next, we need to show that it satisfies the constraint $6)$. The marginal density $\gamma_{(\mathbf{Y}^{(l)})^N}$ can be deduced by the following,
\begin{align*}
\gamma_{(\mathbf{Y}^{(l)})^N}=\dfrac{1}{|C_Y||C_B|}\sum_{y\in C_Y}\sum_{b\in C_B}[\prod_{t=1}^N P_{\mathbf{Y}^{(l)}}(\mathbf{Y}^{(l)}_t|\tilde{y}_t(y,b))]
\end{align*}
We know if $R_{\mathbf{B}^{(l+1)}}+R_{\mathbf{Y}^{(l+1)}} \geq I[\mathbf{Y}^{(l+1)},\mathbf{B}^{(l+1)};\mathbf{Y}^{(l)}|\mathbf{B}^{(l)}]$, then by soft covering lemma \cite{cuff} we have,
\begin{align} \label{eq:6}
\lim_{n\rightarrow\infty} E||\gamma_{(\mathbf{Y}^{(l)})^N}-\prod P_{\mathbf{Y}^{(l)}}||_{TV} = 0
\end{align}

\noindent which shows that $\gamma_{(\mathbf{Y}^{(l)})^N}$ satisfies constraint $6)$. For simplicity of notations we use $\prod P_{\mathbf{Y}^{(l)}}$ instead of $\prod_{t=1}^N P_{\mathbf{Y}^{(l)}}(\mathbf{Y}^{(l)}_t)$, since it can be understood from the context.
Next, let's show that $\gamma_{(\mathbf{Y}^{(l)})^N}$, \textit{nearly} satisfies constraints $2)$ and satisfies $3)$. We need to show that as $N$ grows the synthesized density $\gamma_{(\mathbf{Y}^{(l)})^N,\mathbf{B}^{(l+1)}}$ approaches $\dfrac{1}{|C_B|}\prod P_{\mathbf{Y}^{(l)}}$, in which the latter satisfies both $2)$ and $3)$. In particular, we need to show that the total variation 
$E||\gamma_{(\mathbf{Y}^{(l)})^N,\mathbf{B}^{l+1}}-\dfrac{1}{|C_B|}\prod P_{\mathbf{Y}^{(l)}}||$ vanishes as $N$ grows.
After taking several algebraic steps similar to the ones in \cite{cuff}, we should equivalently show that the following term vanishes, as $N\rightarrow\infty$,
\begin{align} \label{eq:nearly}
\dfrac{1}{|C_B|}\sum_{b\in C_B}E||\gamma_{(\mathbf{Y}^{(l)})^N,\mathbf{B}^{l+1}=\mathbf{b}}-\prod P_{\mathbf{Y}^{(l)}}||_{TV}
\end{align}

Note that given any fixed $b\in C_B$ the number of Gaussian codewords is $|C_Y|=2^{NR_{\mathbf{Y}^{(l+1)}}}$. 
Also, one can check by the signal model defined in \eqref{new_signal} that the statistical properties of the output vector $\mathbf{Y}^{(l)}$ given any fixed sign value $b\in C_B$ does not change. 
Hence, for sufficiently large rates, i.e., $R_{\mathbf{Y}^{(l+1)}}\geq I[\mathbf{Y}^{(l+1)};\mathbf{Y}^{(l)}|\mathbf{B}^{(l)}]$, and by soft covering lemma, the term in the summation in \eqref{eq:nearly} vanishes as $N$ grows.
So overall the term shown in \eqref{eq:nearly} vanishes.
This shows that in fact $\gamma_{(\mathbf{Y}^{(l)})^N}$ \textit{nearly} satisfies the constraints $2)$ and $3)$.
Hence, let's construct another distribution using $\gamma_{(\mathbf{Y}^{(l)})^N,\mathbf{Y}^{(l+1)},\mathbf{B}^{(l+1)}}$. Define,
\begin{align*}
q_{(\mathbf{Y}^{(l)})^N,\mathbf{Y}^{(l+1)},\mathbf{B}^{(l+1)}}=\dfrac{1}{|C_B|}(\prod P_{\mathbf{Y}^{(l)}})\gamma_{\mathbf{Y}^{(l+1)}|(\mathbf{Y}^{(l)})^N,\mathbf{B}^{(l+1)}}
\end{align*}

It is not hard to see that such density satisfies $1)-5)$. 
We only need to show that it satisfies $6)$ as well. We have,
\begin{align}
&||q_{(\mathbf{Y}^{(l)})^N}-\prod P_{\mathbf{Y}^{(l)}}||_{TV}\notag\\
&\leq ||q_{(\mathbf{Y}^{(l)})^N}-\gamma_{(\mathbf{Y}^{(l)})^N}||_{TV} + ||\gamma_{(\mathbf{Y}^{(l)})^N}-\prod P_{\mathbf{Y}^{(l)}}||_{TV}\notag\\
&\leq ||q_{(\mathbf{Y}^{(l)})^N,\mathbf{Y}^{(l+1)},\mathbf{B}^{(l+1)}}-\gamma_{(\mathbf{Y}^{(l)})^N,\mathbf{Y}^{(l+1)},\mathbf{B}^{(l+1)}}||_{TV} + \epsilon_N\label{eq:2nd}\\
&=||q_{(\mathbf{Y}^{(l)})^N,\mathbf{B}^{(l+1)}}-\gamma_{(\mathbf{Y}^{(l)})^N,\mathbf{B}^{(l+1)}}||_{TV} + \epsilon_N\label{eq:4th}\\
&=||\dfrac{1}{|C_B|}(\prod P_{\mathbf{Y}^{(l)}})-\gamma_{(\mathbf{Y}^{(l)})^N,\mathbf{B}^{(l+1)}}||_{TV} +  \epsilon_N \label{eq:condition6}
\end{align}

\noindent where $\epsilon_N=||\gamma_{(\mathbf{Y}^{(l)})^N}-\prod P_{\mathbf{Y}^{(l)}}||_{TV}$.
Both terms in \eqref{eq:condition6} vanish as $N$ grows, due to \eqref{eq:nearly} and \eqref{eq:6}, respectively. Note that, \eqref{eq:2nd} is due to \cite[Lemma V.I]{cuff}. Also, \eqref{eq:4th} is due to \cite[Lemma V.II]{cuff}, by considering the terms $q_{(\mathbf{Y}^{(l)})^N,\mathbf{Y}^{(l+1)},\mathbf{B}^{(l+1)}}$ and $\gamma_{(\mathbf{Y}^{(l)})^N,\mathbf{Y}^{(l+1)},\mathbf{B}^{(l+1)}}$ as the outputs of a unique channel specified by $\gamma_{\mathbf{Y}^{(l+1)}|(\mathbf{Y}^{(l)})^N,\mathbf{B}^{(l+1)}}$, with inputs $p_{(\mathbf{Y}^{(l)})^N,\mathbf{B}^{(l+1)}}$ and $\gamma_{(\mathbf{Y}^{(l)})^N,\mathbf{B}^{(l+1)}}$, respectively.


This completes the achievability proof.

\section{Proof of Lemma \ref{lem:achievability_general2}} \label{app:achievability_general2}

First, we need to change the latent tree structure in a way similar to Figure \ref{fig:internal3}.
We start from the standard latent structure, and at each layer we seek for those latent nodes that are at the same layer and they are neighbors.
For each pair of adjacent nodes, we move the one that is further away from the top layer to a new added layer below the current one.
Hence, make a new layer of latent nodes.
We iterate this step until we reach the bottom layer.
This way, we face different groups of observables being synthesized at different layers.

Define $\mathbf{X}^{(l)}$, $\mathbf{Y}^{(l)}$ and $\mathbf{B}^{(l)}$ as the set of observables, latent nodes and sign variables at layer $l$, respectively.
In this new setting layer $l=0$ defines the observable layer, which only consists of remaining output variables, with no latent nodes.
If the rates at each layer satisfy the inequalities in  \eqref{eq:thm_achievability_basic}, then by Theorem \ref{thm:achievability_basic} we know that as $N$ increases, the simulated density $q_{(\mathbf{X}^{(l)})^N,(\mathbf{Y}^{(l)})^N}$ approaches to the desired density $\prod p_{(\mathbf{X}^{(l)}),(\mathbf{Y}^{(l)})}$.
Suppose the first set of outputs are generated at layer $L'$, then we know $\mathbf{X}=\bigcup_{l=0}^{L'} \mathbf{X}^{(l)}$.
Each observable node ${X}_i^{(l)}$, for $l<L'$ has a latent ancestor at each layer $l<l'\leq L'$. 
We define $\mathbf{Y}'$ as the union of latent nodes containing all those latent ancestors.
Basically, the vector $\mathbf{Y}'$ includes all the latent nodes ${Y}_j^{(l)}$ for $1\leq l\leq L'$.
We define $\mathbf{B}'$, similarly, i.e., those sign inputs related to the nodes in the set $\mathbf{Y}'$.
With slightly abuse of notation, define $\tilde{\mathbf{Y}}=\{\mathbf{Y}',\mathbf{B}'\}$, and $\tilde{\mathbf{Y}}^{(l)}=\{\mathbf{Y}^{(l)},\mathbf{B}^{(l)}\}$, for all possible layers $l$.
The encoding scheme looks exactly as discussed previously, except that this time we need to keep track of corresponding generated outputs at each layer and match them together.
In particular, consider the generated outputs $(\mathbf{X}^{(0)})^N$, which lie at the bottom layer. Each output is generated using a particular input vector $(\mathbf{Y}^{(1)})^N$, which in turn along with other possible outputs $(\mathbf{X}^{(1)})^N$ are generated by a unique input codeword $(\mathbf{Y}^{(2)})^N$ that lie at the second layer.
This procedure moves from the bottom to the top layer, in order to match each generated output at the bottom layer with the correct output vectors at other layers.
Note that the sign information will be automatically taken care of, since similar to the previous cases, at each layer $l+1$ and given each realization of the sign vector $\mathbf{B}^{(l)}=\mathbf{b}^{(l)}$, the input vector $\mathbf{Y}^{(l+1)}$ will become Gaussian.
We only need to show that the synthesize density regarding to such formed joint vectors approaches to the desired output density, as $N$ grows.

By the underlying structure of latent tree, one may factorize the joint density $q_{\mathbf{X}^N,{\tilde{\mathbf{Y}}}^N}=q_{(\mathbf{X}^{(L')})^N,(\tilde{\mathbf{Y}}^{(L')})^N}\prod_{l=0}^{L'-1}q_{(\mathbf{X}^{(l)})^N|(\tilde{\mathbf{Y}}^{(l+1)})^N}$.
Note that the desired joint density $p_{\mathbf{X},\tilde{\mathbf{Y}}}$ also induces the same latent Gaussian tree, hence, we may write, $p_{\mathbf{X}^N,{\tilde{\mathbf{Y}}}^N}=p_{(\mathbf{X}^{(L')})^N,(\tilde{\mathbf{Y}}^{(L')})^N}\prod_{l=0}^{L'-1}p_{(\mathbf{X}^{(l)})^N|(\tilde{\mathbf{Y}}^{(l+1)})^N}$.
However, by our encoding scheme shown in Figure \ref{fig:encoding_basic}, one may argue that $\prod_{l=0}^{L'-1}q_{(\mathbf{X}^{(l)})^N|(\tilde{\mathbf{Y}}^{(l+1)})^N}=\prod_{l=0}^{L'-1}p_{(\mathbf{X}^{(l)})^N|(\tilde{\mathbf{Y}}^{(l+1)})^N}=\prod_{l=0}^{L'-1}\prod p_{\mathbf{X}^{(l)}_t|\tilde{\mathbf{Y}}^{(l+1)}_t}$.
By summing out $(\mathbf{B}^{(L')})^N$ from both densities $p_{\mathbf{X}^N,{\tilde{\mathbf{Y}}}^N}$ and $q_{\mathbf{X}^N,{\tilde{\mathbf{Y}}}^N}$, we may replace $p_{(\mathbf{X}^{(L')})^N,(\tilde{\mathbf{Y}}^{(L')})^N}$ with $p_{(\mathbf{X}^{(L')})^N,({\mathbf{Y}}^{(L')})^N}$ and $q_{(\mathbf{X}^{(L')})^N,(\tilde{\mathbf{Y}}^{(L')})^N}$ with $q_{(\mathbf{X}^{(L')})^N,({\mathbf{Y}}^{(L')})^N}$, since only these terms in the equations depend on the sign vector at layer $L'$, i.e., $(\mathbf{B}^{(L')})^N$.
Now, by previous arguments for the synthesized and desired density at layer $L'$, we know that the total variation distance $||q_{(\mathbf{X}^{(L')})^N,(\mathbf{Y}^{(L')})^N}-\prod p_{\mathbf{X}^{(L')}_t,\mathbf{Y}^{(L')}_t}||_{TV}$ goes to zero as $N$ grows.
Hence, one may simply deduce that $||q_{\mathbf{X}^N,{\tilde{\mathbf{Y}}}^N/(\mathbf{B}^{(L')})^N}-\prod p_{\mathbf{X}_t,{\tilde{\mathbf{Y}}}_t/\mathbf{B}^{(L')}_t}||_{TV}=||(q_{(\mathbf{X}^{(L')})^N,(\mathbf{Y}^{(L')})^N}-\prod p_{\mathbf{X}^{(L')}_t,\mathbf{Y}^{(L')}_t})\prod_{l=0}^{L'-1}\prod p_{\mathbf{X}^{(l)}_t|\tilde{\mathbf{Y}}^{(l+1)}_t}||_{TV}$ goes to zero as $N$ grows.
Due to \cite[Lemma V.I]{cuff}, we know $||q_{\mathbf{X}^N}-\prod p_{\mathbf{X}_t}||_{TV}\leq ||q_{\mathbf{X}^N,{\tilde{\mathbf{Y}}}^N/(\mathbf{B}^{(L')})^N}-\prod p_{\mathbf{X}_t,{\tilde{\mathbf{Y}}}_t/\mathbf{B}^{(L')}_t}||_{TV}<\epsilon$, and as $N$ grows.
This completes the proof.

\end{document}